\documentclass[a4paper,10pt]{article}

\usepackage{amsthm}
\usepackage{amsmath}
\usepackage{amssymb}

\usepackage{tikz}
\usepackage[
]{hyperref}
\usepackage{enumerate}
\usepackage{setspace}

\newcommand{\Con}{\ensuremath{\mathcal{C}}}

\newcommand{\Cinf}{\ensuremath{\mathcal{C}^\infty}}

\newcommand{\D}{\ensuremath{{\mathcal D}}}

\newcommand{\loc}{\ensuremath{\text{loc}}}

\newcommand{\mb}[1]{\ensuremath{\mathbb{#1}}}
\newcommand{\N}{\mb{N}}

\newcommand{\R}{\mb{R}}

\newcommand{\cl}[1]{\ensuremath{[#1]}}

\newcommand{\G}{\ensuremath{{\mathcal G}}}

\newcommand{\EM}{\ensuremath{{\mathcal E}_{\mathrm{M}}}}

\newcommand{\NN}{\ensuremath{{\mathcal N}}}

\renewcommand{\d}{\ensuremath{\partial}}

\newcommand{\pder}[2]{\frac{\d#1}{\d#2}}

\newcommand{\grad}{\ensuremath{\mbox{\rm grad}\,}}

\newfont{\bl}{msbm10 scaled \magstep2}

\newcommand{\beq}{\begin{equation}}
\newcommand{\eeq}{\end{equation}}

\newcommand{\emb}{\hookrightarrow}

\newcommand{\col}{\colon}

\newcommand{\dis}[2]{\langle #1 , #2 \rangle}
\newcommand{\notmid}{\mid\kern-0.5em\not\kern0.5em}

\newcommand{\norm}[2]{{\| #1 \|}_{#2}}

\newcommand{\eps}{\varepsilon}

\newcommand{\vphi}{\varphi}

\newcommand{\la}{\lambda}

\newenvironment{pr}{\begin{proof}[\textbf{Proof:}] \ }{\end{proof}}
\newtheorem{thm}{Theorem}[section]
\newtheorem{lem}[thm]{Lemma}
\newtheorem{prop}[thm]{Proposition}

\newtheorem{cor}[thm]{Corollary}
\newtheorem{defi}[thm]{Definition}
\theoremstyle{definition}
\newtheorem{rem}[thm]{Remark}

\newcommand{\Fl}[2]{\text{Fl}_{#1}^{#2}}
\newcommand{\tla}{\tau_\lambda}
\newcommand{\ep}{\epsilon}
\newcommand{\subcomp}{\Subset}
\newcommand{\Ric}{\text{Ric}}

\numberwithin{equation}{section}

\title{The globally hyperbolic metric splitting for non-smooth wave-type space-times}
\author{G\"unther H\"ormann \&	 Clemens S\"amann}

\begin{document}

\maketitle

\begin{abstract}
We investigate a generalization of the so-called metric splitting of globally hyperbolic space-times to non-smooth
Lorentzian manifolds and show the existence of this metric splitting for a class of wave-type space-times. The approach used
is based on smooth approximations of non-smooth space-times by families (or sequences) of globally hyperbolic space-times.
In the same setting we indicate as an application the extension of a previous result on the Cauchy problem for the wave
equation.
\end{abstract}


\section{Introduction}
We investigate causal properties, especially global hyperbolicity, of wave-type space-times. The relevance of global
hyperbolicity in general relativity is due to its role as the strongest established causality condition, in particular in the
context of Cauchy problems and singularity theorems. Several equivalent conditions of global hyperbolicity have been
investigated and developed, one of the first was existence of a Cauchy hypersurface and the most recent breakthrough was the
proof the so-called \emph{metric splitting} (cf.\ \cite{BS:05}; see also the discussion and Theorem \ref{thm-bs} in Section
\ref{sec-sm}). To initiate research for an extension of global hyperbolicity to the situation of non-smooth space-times this
article aims at providing a case study, thereby also describing the explicit form of the metric splitting in the smooth case.
For an overview of wider applications in general relativity of non-smooth Lorentzian metrics with techniques similar to the
methods used here we refer to \cite{SV:06}.

By wave-type space-times we mean a generalization of plane waves, the so-called $N$-fronted waves with parallel rays (NPWs) or
general plane fronted waves (PFWs). These space-times are given as a product $M=N\times \R^2$, with metric
\begin{equation}\label{eq-int-met}
 l\ =\ \pi^*(h) + 2dudv - a(.,u)du^2\ ,
\end{equation}
where $h$ denotes the metric of an arbitrary connected Riemannian manifold $(N,h)$, $\pi\colon M\rightarrow N$ is the
projection ($\pi^*(h)$ denotes the pullback under the projection of $h$ to $M$) and $u,v$ are global
null-coordinates on the two-dimensional Minkowski space $\R^2_1$. Moreover $a\colon N\times\R\rightarrow\R$ is the so-called
profile function, which we allow to be non-smooth. Locally in coordinates $x^1,\ldots, x^n$ on $N$ at $(x,u,v)\in M$
the metric $l$ can be written as
\begin{equation*}
 l_{(x,u,v)}\ =\ \sum_{i,j=1}^n h_{ij} dx^i dx^j + 2dudv - a(x,u)du^2\ ,
\end{equation*}
where $h_{ij}$ denote the metric coefficients of $h$ with respect to $x^1,\ldots,x^n$.

NPWs were introduced by Brinkmann in the context of conformal mappings of Einstein spaces (\cite{Brinkmann:25}). Recently
their geometric properties and causal structure were studied in \cite{CFS:03, FS:03, CFS:04, FS:06} (under the notion of
general plane fronted waves - PFW). Due to the geometric interpretation of $N$ as the wave surface of
these waves (cf.\ \cite{SS:12}), it seems more natural to call them $N$-fronted waves, rather than plane-fronted
waves. Note that plane-fronted waves with parallel rays (pp-waves) (cf.\ \cite[Ch.17]{GP:09}) are a special case of NPWs. In
this case $N=\R^2$ with the Euclidean metric. 

\bigskip
It turns out (in the classical setting where the metric is smooth) that the behavior of $a$ at spatial infinity, i.e., for
``large $x$'' is decisive for many of the global properties of NPWs. In order to formulate precise statements denote by $d^h$
the Riemannian distance function on $(N,h)$ and recall that $a$ is said to behave \emph{subquadratically at spatial
infinity}, if there exist a point $\bar x\in N$, continuous non-negative functions $R_1$, $R_2\colon\R \rightarrow
(0,\infty)$ and a continuous function $p\colon\R \rightarrow (0,2)$ such that for all $(x,u)\in N\times\R$
\begin{equation}\label{eq-sq} 
a(x,u)\leq R_1(u)d^h(x,\bar x)^{\,p(u)}+R_2(u)\ .
\end{equation} 
Similarly we say that $a$ behaves at most quadratically if $p\leq 2$.
In \cite{FS:03} it has been shown that the causality of NPWs depends crucially on the exponent $p$ in (\ref{eq-sq}), with
$p=2$ being the critical case, which includes classical plane waves that are known to be strongly causal but not globally
hyperbolic (cf.\ \cite{Penrose:65}). In particular, NPWs are causal but not necessarily distinguishing, they are strongly
causal if $a$ behaves at most quadratically at spatial infinity and they are globally hyperbolic if $a$ is subquadratic and
$N$ is complete. Similarly the global behavior of geodesics in NPWs is governed by the behavior of $a$ at spatial infinity.
From the explicit form of the geodesic equations it follows (\cite[Thm.\ 3.2]{CFS:03}) that a NPW is geodesically complete if
and only if $N$ is complete and
\begin{equation}\label{eq-intro-traj}
 D^N_{\dot \xi}\dot \xi = \frac{1}{2} \nabla_x a(\xi, \alpha)
\end{equation}
has complete trajectories for all $\alpha\in\R$, i.e., the solutions of \eqref{eq-intro-traj} can be defined on all of $\R$. Here $D^{N}_{\dot \xi}$
is the induced covariant derivative on $N$ and $\nabla_x$ denotes the metric gradient with respect to $h$. Applying classical
results on complete vector fields (e.g.\ \cite[Thm.\ 3.7.15]{AMR:88}) completeness of $M$ follows for autonomous~$a$ (i.e.,
independent of $u$) in case $-a$ grows at most quadratically at spatial infinity.

\bigskip

When discussing the case of non-smooth profile function $a$ we will also employ the nonlinear theory of generalized functions in the sense of Colombeau, standard references are \cite{Colombeau:84,Colombeau:85,O:92,GKOS:01}. Our framework is 
the so-called special Colombeau algebra $\G$ (denoted by $\G^s$ in \cite{GKOS:01}) and we briefly recall the basic constructions. Let $M$ be a smooth manifold. \emph{Colombeau generalized functions} on
$M$ are defined as equivalence classes $u = \cl{(u_\eps)_\eps}$ of nets of
smooth functions $u_\eps\in\Cinf(M)$ (\emph{regularizations}) subjected to
asymptotic norm conditions with respect to $\eps\in (0,1]$ for their
derivatives on compact sets. More precisely, we have
\begin{itemize}
 \item moderate nets $\EM(M)$: $(u_\eps)_\eps\in\Cinf(M)^{(0,1]}$ such that
 for any compact subset $K \subseteq M$, $l \in \N_0$, and vector fields $X_1, \ldots,X_l$ on $M$ there exists $p \in \R$ such that
 $$
    \norm{X_l \cdots X_1 u_\eps}{L^\infty(K)} = O(\eps^{-p}) \qquad (\eps \to 0)\ ;
 $$
  \item negligible nets $\NN(M)$: $(u_\eps)_\eps\in \EM(M)$ such that for every compact subset $K \subseteq M$ and $q\in\R$ an
 estimate $\norm{u_\eps}{L^\infty(K)} = O(\eps^{q})$ ($\eps \to 0$) holds;
 \item $\EM(M)$ is a differential algebra with operations defined at fixed $\eps$,
 $\NN(M)$ is an ideal, and $\G(M) := \EM(M) / \NN(M)$ is the (special)
 \emph{Colombeau algebra};
 \item there are embeddings, $\Cinf(M)\emb \G(M)$ as subalgebra and
 $\D'(M) \emb \G(M)$ as linear space.
 \item For the discussion of mappings into manifolds and compositions the notion of c-boundedness is crucial (cf.\ \cite[Def.
1.2.7]{GKOS:01}). A moderate net of maps is called \emph{c-bounded} if the images of any compact set are contained in a fixed
compact set in the target space for small $\ep$. 
 \item By $\G[M,N]$ we denote c-bounded generalized maps from $M$ to $N$.
\end{itemize}

\bigskip
The outline of the paper is as follows: As a preparation, in Section \ref{sec-sm} we suppose that $a$ is smooth, hence we are able to employ methods of
smooth differential geometry. Then in the third section we apply these results to nets of smooth functions thus entering the
framework of Colombeau generalized functions. Finally, in the fourth section we use these methods and results in
approximating non-smooth profile functions $a\colon N\times\R\rightarrow \R$ and indicate applications to wave-equations on
space-times with low regularity in the metric.


\section{The smooth metric splitting}\label{sec-sm}
Let $(N,h)$ be an $n$-dimensional smooth, connected Riemannian manifold and $M=N\times\R^2$. Let $0 \leq a\in\Cinf(N\times\R)$. The metric $l$ on $M$ is given by
\begin{equation*}
 l:=\pi^*(h) + 2 du dv - a(.,u)du^2\ ,
\end{equation*}
where $\pi^*(h)$ denotes the pullback under the projection of $h$ to $M$. We write $l$ in coordinates, with $h_x$ denoting the local matrix
representation of $h$ at $x\in N$, as 
\begin{equation}\label{eq-l-mat}
 l_{(x,u,v)} =  \begin{pmatrix}
		 h_x & 0     &0\\
                 0 & -a(x,u) & 1\\
		 0 &  1    & 0
              \end{pmatrix}
 \quad \text{and }\ 
 l_{(x,u,v)}^{-1} = \begin{pmatrix}
                    h^{-1}_x & 0     &0\\
		    0 & 0 & 1\\
		    0 & 1 & a(x,u)
                  \end{pmatrix}\ .
\end{equation}

The eigenvalues of $l$ at $(x,u,v)$ consist of
\begin{equation}\label{eq-cc-ev-gen-case}
 \mu_2:= \frac{-a(x,u)}{2} - \sqrt{\frac{a(x,u)^2}{4}+1} < 0 < \frac{-a(x,u)}{2} + \sqrt{\frac{a(x,u)^2}{4}+1} =: \mu_1
\end{equation}
and the positive eigenvalues $\nu_1, \ldots,\nu_n$ of $h_x$, as can be seen from $\det(l-\mu I_{n+2}) = \det(h_x-\mu I_n) \det\begin{pmatrix}
                -a(x,u)-\mu & 1\\
		 1          & -\mu
              \end{pmatrix}$, since the matrix $l-\mu I_{n+2}$ is in block-diagonal form. (Here $I_k$ denotes the
$k$-dimensional identity matrix.) Since $\mu_2 < 0 < \mu_1,\nu_1,\ldots,\nu_n$, we see that  $l$ has index $1$ and is
non-degenerate, hence $(M,l)$ is a Lorentzian manifold. Moreover, note that $\det(l)=-\det(h)$ does not depend on the profile function $a$.

\subsection*{Causal curves and causality}
\addcontentsline{toc}{subsection}{Causal curves and Causality}
As in \cite[p.\ 83]{FS:06} the time-orientation on $(M,l)$ is chosen such that $\d_v$ is past-directed. Let $J\subseteq \R$
be an interval and let $\gamma=(\xi, \alpha,\beta) \colon J \rightarrow M$ be a causal curve, i.e., for all $s\in J$
\begin{equation}\label{eq-ccc-cc-full}
 0\geq l(\dot\gamma(s),\dot\gamma(s)) = h_{\xi(s)}(\dot\xi(s),\dot\xi(s)) + 2
\dot\alpha(s)\dot\beta(s)-a(\xi(s),\alpha(s))\dot\alpha(s)^2\ .
\end{equation}
By positive definiteness of $h$ we have
\begin{equation}\label{eq-ccc-cc}
 0\geq l(\dot\gamma(s),\dot\gamma(s)) - h(\dot\xi,\dot\xi) = 2
\dot\alpha(s)\dot\beta(s)-a(\xi(s),\alpha(s))\dot\alpha(s)^2\quad \forall s\in J\ .
\end{equation}
If $\gamma$ is timelike, i.e., $l(\dot\gamma,\dot\gamma)<0$ and future directed, then $\dot\alpha>0$. In fact,
\eqref{eq-ccc-cc} then implies $\dot\alpha\neq0$ on $J$ and since $\d_v$ is past directed
we obtain that $0<l(\d_v,\dot\gamma)=\dot\alpha$. 

\bigskip
From now on we assume that $a$ is bounded, i.e., there exists $\lambda_0>0$ such that $0\leq a(x,u)< 2 \lambda_0$ for all $(x,u)\in
N\times \R$. In particular $a$ is subquadratic (cf.\ \eqref{eq-sq}).

\begin{figure}[!h]
\begin{tikzpicture}[scale=3.2] 
\draw [<->] (1,0)--(-1,0);
\draw [<->] (0,-1)--(0,1);
\node [below right] at (1,0) {$v$};
\node [left] at (0,1) {$u$};

\draw [very thick, ->] (0,0)--(.2,0) node [below] {$\d_v$};
\draw [thick, blue] (-.45, -1)--(.45, 1) node [left, black] {$S$};
\path [fill=cyan, opacity=0.3] (0, 0)--(.2, 1)--(-1,1)--(-1,0)--(0,0);
\path [fill=lightgray, opacity=0.3] (0, 0)--(-.2, -1)--(1,-1)--(1,0)--(0,0);
\node at (.5, .3) {space};
\node at (-.5, -.7) {space};
\node at (-.5, .3) {future};
\node at (.6, -.7) {past};

\draw[dashed] (-.2, -1)--(.2,1);
\node at (-.55,.75) [right] {slope $\frac{a(u_0)}{2}$};
\draw[red, <->] (0,.8)--(.165,.8);
\end{tikzpicture}
\centering
\caption{Light cones of $l$ in the two-dimensional case at the point $(u_0,v_0)\in\R^2$}
\end{figure}
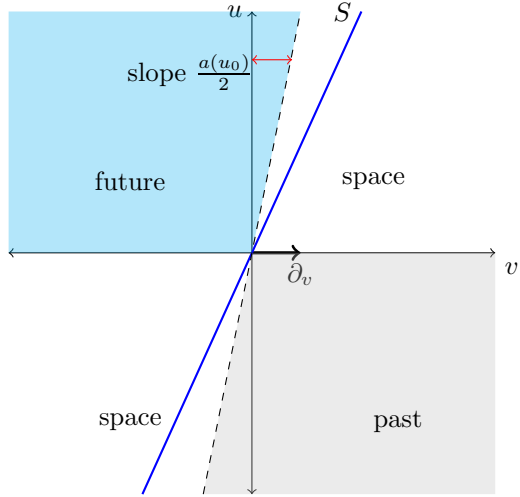
Figure 1 indicates the behavior of the lightcones depending on $a(u_0)$. In this two-dimensional setting we
see that $a(u_0)$ gives the slope of the lightlike line $v=\frac{a(u_0)}{2}u$ and the spacelike line $S=\{(u,\lambda_0 u):
u\in\R\}$ indicates the ``maximal'' slope. Therefore one can see that the lightcone varies between the Minkowski
lightcone ($a(u)=0$) and the spacelike line $S$ ($a(u)=\lambda_0$), where the boundary case $a(u)=0$ is included and
$a(u)=2\lambda_0$ is excluded.

\bigskip
\textbf{Fact 1:}\phantomsection\label{thm-cc-sc}
 The Lorentzian manifold $(M,l)$ is strongly causal.\\
This follows from \cite[Thm.\ 3.1]{FS:03} since
$a$ is obviously subquadratic. 

\bigskip
\textbf{Fact 2:}\phantomsection\label{thm-gh}
 If the Riemannian manifold $(N,h)$ is complete, then the Lorentzian manifold $(M,l)$ is globally hyperbolic. \\
 This is a consequence of the (obvious) subquadratic behavior of $a$ and is shown in \cite[Thm.\ 4.1]{FS:03}.

\bigskip
From now on let the Riemannian manifold $(N,h)$ be complete.

\subsection*{Geodesics}
\addcontentsline{toc}{subsection}{Geodesics}
The only non-vanishing Christoffel symbols (cf.\ \cite{CFS:03}) are
\begin{align*}
 &\Gamma^k_{ij} = \Gamma^{k(N)}_{ij}  &(i,j,k=1,\ldots,n)\ ,\\
 &\Gamma^v_{uj}=\Gamma^v_{ju}=-\frac{1}{2}\pder{a}{x^j}  &(j=1,\ldots,n)\ ,\\
 &\Gamma^k_{uu} = \frac{1}{2}\sum_{i=1}^n h^{ki}\pder{a}{x^i} &(k=1,\ldots,n)\ ,\\
 &\Gamma^v_{uu} = -\frac{1}{2}\pder{a}{u}\ ,
\end{align*}
where $x^1,\ldots, x^n$ is a coordinate system on $N$. For a curve $\gamma=(\xi,\alpha,\beta)\colon J\rightarrow M$ the
geodesic
equations read
\begin{align} 
 &D^N_{\dot \xi}\dot \xi = \frac{1}{2} \nabla_x a(\xi, \alpha)\ \label{eq-geo-x-n},\\
 &\ddot \alpha = 0 \label{eq-geo-a-n}\ ,\\
 &\ddot \beta = \sum_{j=1}^n \pder{a}{x^j}(\xi,\alpha)\dot \xi^j + \frac{1}{2} \dot\alpha^2 \pder{a}{u}(\xi, \alpha)\label{eq-geo-b-n}\ ,
\end{align}
where $D^N$ denotes the covariant derivative on $N$ with respect to $h$ and $\nabla_x$ denotes the gradient with respect to
$h$.

\subsection*{Time functions and Cauchy hypersurfaces}
\addcontentsline{toc}{subsection}{Time functions and Cauchy hypersurfaces}
Since $(M,l)$ is globally hyperbolic (by assuming that $(N,h)$ is complete and Fact \ref{thm-gh}) we know there exist
time functions (cf.\ for example \cite[p.\ 64]{BEE:96}), i.e., continuous functions $f\colon M\rightarrow \R$ such that $f$
is strictly increasing along future directed causal curves. In fact we even have a so-called \emph{temporal} function, i.e.,
$f$ is smooth and has past directed timelike gradient (\cite[Def.\ 3.48]{MS:08}). Recall that we assumed $a$ to be
bounded.

\begin{prop}\label{prop-tla}
 Let $\lambda>\|a\|_\infty$ and define $\tla\colon M\rightarrow \R$ by $\tla(x,u,v):= \lambda u - v$. Then $\tla$ is a
temporal function for $(M,l)$, hence also
a time function.
\end{prop}
\begin{pr}
 The gradient of $\tla$ is $\grad{\tla} = -\d_u + (\lambda - a(x,u))\d_v$ at $(x,u,v)\in M$ and it is everywhere
timelike since $l(\grad{\tla},\grad{\tla}) = a - 2\lambda < -\lambda < 0$. Furthermore $\grad{\tla}$ is past directed:
$l(\d_v,\grad{\tla})=-1<0$. 
\end{pr}

So, for every $\lambda > \|a\|_\infty$ we obtain a time function $\tla$. Is it also a Cauchy time function (cf.\
\cite[p.\ 65]{BEE:96})? To be a Cauchy time function it has to satisfy that
\begin{equation*}
 S^k_\lambda:= \tla^{-1}(\{k\}) = \{(x,u,\lambda u - k): x\in N, u\in\R\}
\end{equation*}
is a Cauchy hypersurface (every inextendible timelike curve meets the surface exactly once) for every $k\in\R$.

\begin{thm}\label{thm-chs}
 If equation \eqref{eq-geo-x-n} has complete trajectories or $M=\R^2$, then for every $k\in\R$ the
set
$S^k_\lambda$ is a Cauchy hypersurface in $(M,l)$. 
\end{thm}
\begin{pr}
  Let $k\in\R$, then by \cite[Prop.\ 4.17]{ONeill:83} $S^k_\lambda$ is a semi-Riemannian hypersurface since
$l(\grad(\tla),\grad(\tla))<0$ everywhere and $S^k_\lambda=\tla^{-1}(\{k\})$ (and it is obviously non-empty). The
hypersurface $S^k_\lambda$ is spacelike: Let $p=(x,u, \lambda u - k)\in S^k_\lambda$, then any $\eta\in
T_p S^k_\lambda$ is of the form $\eta=(\xi, \alpha, \lambda \alpha)$, where $\xi\in T_x N$, $\alpha\in\R$.
If $\eta\neq 0$ then $l(\eta,\eta)=h(\xi,\xi)+2\lambda\alpha^2 - a(x,u)\alpha^2>h(\xi,\xi)+\lambda\alpha^2>0$.
Since $S^k_\lambda$ is closed (as continuous preimage of the closed set $\{k\}$) and since
$S^k_\lambda$ separates $M$ ($M\backslash S^k_\lambda$ is obviously not connected) \cite[Lem.\ 14.45(2)]{ONeill:83}
shows that $S^k_\lambda$ is achronal. An achronal spacelike hypersurface is acausal (\cite[Lem.\ 14.42]{ONeill:83}),
so by \cite[Cor.\ 14.54]{ONeill:83} it suffices to show that every inextendible null geodesic meets $S^k_\lambda$.

\bigskip
First we prove the assertion in the two-dimensional case, i.e., $M=\R^2$: let $\gamma=(\alpha,\beta)\colon (A,B)\rightarrow
M$ be an inextendible null geodesic. Then either $\dot\alpha(t)=0$ for all $t\in(A,B)$ or $\dot\beta(t) =
\frac{1}{2}a(\alpha(t))\dot\alpha(t)\ \forall t\in(A,B)$, where $\alpha(t)=a_1 t + a_0$ by \eqref{eq-ccc-cc} (with
$l(\dot\gamma,\dot\gamma)=0$) and \eqref{eq-geo-b-n} in this simplified situation with $x$-independent $a$. We have to show that there is a $t^*\in(A,B)$ such that $\beta(t^*)=\lambda\alpha(t^*)-k$.

Case $\dot \alpha = 0$: We have $a_1=0$, hence $\alpha$ is constant. 

First, if $\beta(t)< \lambda \alpha(t) - k = \lambda a_0 - k\ \forall t\in(A,B)$, we conclude that
$\beta$ is bounded from above. Moreover since $\dot\alpha=0$ we know that $\dot\beta\neq 0$ because otherwise the tangent
vector $\dot\gamma$ would be spacelike. Therefore either $\dot\beta<0$ or $\dot\beta>0$ on $(A,B)$. But in both cases we get
a contradiction since $\dot\beta<0$ implies that $\lim_{t\searrow A}\beta(t)$ exists, which contradicts the inextendibility of $\gamma$. Similarly, if $\dot\beta>0$ we conclude that $\lim_{t\nearrow B}\beta(t)$ exists. Analogously we
can handle the second sub-case where $\beta(t)>\lambda\alpha(t)-k = \lambda a_0 -k\ \forall t\in(A,B)$.

Case $\dot\beta(t) = \frac{1}{2}a(\alpha(t))\dot\alpha(t)$: We have that $a_1\neq 0$, since otherwise $\dot\gamma=0$ would be spacelike. Therefore $\dot\beta(t) = \frac{1}{2}a(\alpha(t))\dot\alpha(t) = \frac{1}{2} a(a_1 t + a_0)a_1 \neq 0\ \forall
t\in(A,B)$.  Now as above we
assume that $\beta(t)>\lambda\alpha(t)-k\ \forall t\in(A,B)$ and without loss of generality also that $\dot\beta>0$ (the
other case is analogous). Then we integrate to obtain
\begin{align*}
 \beta(t)-\beta(t_0) &= \int_{t_0}^t\dot\beta(s)ds = \frac{1}{2}\int_{t_0}^t a(\alpha(s))\dot\alpha(s)ds =
\frac{1}{2}\int_{\alpha(t_0)}^{\alpha(t)} a(s) ds \\
& < \frac{\lambda}{2}(\alpha(t)-\alpha(t_0)) < \frac{\beta(t)}{2} + \frac{k-\lambda\alpha(t_0)}{2}\ ,
\end{align*}
where we used the hypotheses of this sub-case. From this inequality we conclude that $\beta(t)< 2\beta(t_0)+
k-\lambda\alpha(t_0)$ for all $t>t_0$, hence $\beta$ is bounded from above. This yields (as in the first case) the existence
of $\lim_{t\nearrow B}\beta(t)$ and, since $\alpha$ is just a straight line, this contradicts the inextendibility of
$\gamma$ in the case $B$ is finite and in the case $B=+\infty$ the fact that $\beta(t) > \lambda
\alpha(t)-k\ \forall t\in (A,B)$. Analogously one can show that the assumption $\beta(t)<\lambda\alpha(t)-k\ \forall
t\in(A,B)$
cannot hold.

In summary we get that there is a $t^*\in(A,B)$ with $\beta(t^*)=\lambda\alpha(t^*)-k$, hence $\gamma(t^*)\in
S^k_\lambda$, and so $S^k_\lambda$ is a Cauchy hypersurface.

\bigskip
Now for the case $n = \dim N >0$, where we assume that equation \eqref{eq-geo-x-n} has complete trajectories. Let
$\gamma=(\xi,\alpha,\beta)\colon (A,B)\rightarrow M$ be an inextendible null geodesic. Since it has complete trajectories we
can
assume that $(A,B)=\R$. Our first observation is that, since $\gamma$ is null, we get from \eqref{eq-ccc-cc-full}
\begin{equation}\label{eq-chs-null}
 0\leq h(\dot \xi, \dot \xi) = (a(\xi,\alpha)\dot \alpha - 2\dot\beta)\dot\alpha\ .
\end{equation}
Moreover from \eqref{eq-geo-a-n} we know that $\dot\alpha$ is constant, so we get that in the case $\dot\alpha=0$, the
general case reduces to the two-dimensional case since, now $\xi$ is constant by equation \eqref{eq-chs-null}. In the case
$\dot\alpha\neq0$ we can without loss of generality assume that $\dot\alpha>0$ and write $\alpha(t)=a_1 t + a_0\ (t\in\R)$
with $a_0, a_1\in \R, a_1>0$. The case $a_1<0$ is analogous. Rearranging \eqref{eq-chs-null} we get
\begin{equation}\label{eq-chs-b}
\dot\beta = \frac{1}{2} \left(a(\xi,\alpha)a_1 - \frac{1}{a_1}h(\dot\xi,\dot\xi)\right)\ .  
\end{equation}
Integrating \eqref{eq-chs-b} we get that
\begin{equation*}
 \beta(t)=b_0 + \frac{a_1}{2} \int_0^t a(\xi(s),a_1 s + a_0) ds - \frac{1}{2 a_1}\int_0^t h(\dot\xi(s),\dot\xi(s))ds\quad
(t\in\R)\ .
\end{equation*}
We conclude that $\beta(t) \leq b_0 + \frac{a_1}{2}\lambda t$, so since $\lambda\alpha-k$ has slope $\lambda a_1$,
the curves will eventually meet if we start out with $\lambda \alpha(0) - k = \lambda a_0 - k < b_0 = \beta(0)$. Analogously
if we start out with $\lambda a_0 - k > b_0$, we get a point of intersection.

In summary we get a point $t^*\in\R$ such that $\lambda \alpha(t^*) - k = \beta(t^*)$, therefore
$\gamma(t^*)=(\xi(t^*),\alpha(t^*),\lambda \alpha(t^*)+k)\in S^k_\lambda$. 
\end{pr}
In particular, $\tla$ is a Cauchy time function under the hypotheses of Theorem \ref{thm-chs}.

\bigskip
Candela, Romero, and S\'anchez give sufficient conditions on $a$ ensuring completeness of the trajectories of \eqref{eq-geo-x-n} in \cite{CRS:12}
(especially Theorem 2 and Subsection 3.2).

%
%

\subsection*{Metric splitting} \label{sub-ms}
\addcontentsline{toc}{subsection}{Metric splitting}
The normalized gradient of $\tla$ is 
\begin{equation*}
Y:=\frac{\grad{\tla}}{l(\grad{\tla},\grad{\tla})} = \frac{1}{2\lambda-a(x,u)}\d_u + \frac{a(x,u)-\lambda}{2\lambda -
a(x,u)}\d_v\quad \bigl((x,u,v)\in M\bigr)\ .
\end{equation*}
Calculating the flow as explicitly as possible will be crucial when we apply these results to the non-smooth
case. Solving $\dot\gamma(t)=Y(\gamma(t))$ is equivalent to solving the system
\begin{align}
 \dot\xi(t) &= 0\notag\ ,\\
 \dot\alpha(t) &= \frac{1}{2\lambda - a(\xi(t),\alpha(t))}\ ,\label{eq-fl-a}\\
 \dot\beta(t) &= \frac{a(\xi(t),\alpha(t))-\lambda}{2\lambda - a(\xi(t),\alpha(t))}\ ,\label{eq-fl-b}
\end{align}
where $\gamma=(\xi,\alpha,\beta)$. From our assumptions on $a$ and $\lambda$ we see that for every flow 
$\gamma$ of $Y$ we have that $\dot\alpha>0$ and $\dot\beta<0$. Furthermore from the structure of the equations
\eqref{eq-fl-a}, \eqref{eq-fl-b} it follows that we only have to solve for $\alpha$, since then $\beta$ can be found by
integration.

\begin{thm}(The flow of the normalized gradient of $\tla$)\label{thm-flow-sm}
 The flow  of $Y$ is given by
\begin{equation*}
 \Fl{t}{Y}(x,u,v)= \begin{pmatrix} x\\K_x(t,u)\\ v-\lambda u - t + \lambda K_x(t,u) \end{pmatrix}\quad \bigl((x,u,v)\in M,
t\in\R\bigr)\ ,
\end{equation*}
where $K_x(t,u):= F_x^{-1}(t+F_x(u))$ and $F_x(u):=\int_0^u(2\lambda - a(x,s))ds$.
\end{thm}
\begin{pr}
First we observe that $\xi$ is constant, so we can fix $x\in N$. Since $F_x'(u)=2\lambda - a(x,u) > 0$, we conclude that
$F_x$ is strictly monotonically increasing and $F_x$ is also surjective ($\lambda u < F_x(u) \leq 2\lambda u$). So
$F_x\colon\R \rightarrow \R$ is bijective, hence $F_x^{-1}\colon\R \rightarrow \R$ exists and we can define $K_x\colon\R^2
\rightarrow \R$ as
above. Clearly, $F_x$ and $F_x^{-1}$ are smooth as functions of $u$, since $a$ is smooth. Moreover from the defining equation
$F(x,K(x,t,u))=t+F(x,u)$ and the implicit function theorem we obtain smoothness of $F$ and $K$ (compare the argument in
Section \ref{sec-nsm}, around equation \eqref{eq-impl-fct}).

To solve \eqref{eq-fl-a} we integrate the equation to obtain
\begin{equation*}
 t = \int_0^t(2\lambda - a(x,\alpha(r)))\dot\alpha(r)dr = \int_{\alpha(0)}^{\alpha(t)}(2\lambda-a(x,s))ds =
F_x(\alpha(t))-F_x(\alpha(0))\ ,
\end{equation*}
where we substituted $s=\alpha(r)$ (note that $\alpha$ is monotonically increasing). This yields that
$\alpha(t)=F_x^{-1}(t+F_x(\alpha(0))) = K_x(t,\alpha(0))$.

Rewriting \eqref{eq-fl-b} as $\dot\beta(t)= -1 + \frac{\lambda}{F_x'(\alpha(t))}$ and then integrating we get
\begin{align*}
 \beta(t) &= \beta(0)-t + \lambda \int_0^t\frac{1}{F_x'(\alpha(r))}dr \\
 &= \beta(0)-t + \lambda\int_0^t\frac{1}{F_x'(F_x^{-1}(r + F_x(\alpha(0))))}dr \\
 &=\beta(0)-t+\lambda\int_0^t(F_x^{-1})'(r+F_x(\alpha(0)))dr\\
 &= \beta(0) - t + \lambda(F_x^{-1}(t +F_x(\alpha(0)))-F_x^{-1}(F_x(\alpha(0))))\\
 &= \beta(0)-\lambda \alpha(0) - t +\lambda K_x(t,\alpha(0))\ .
\end{align*}
\end{pr}
We restrict $\Fl{ }{Y}$ to $S:=S^0_\lambda=\tla(\{0\}) = \{(x,u,\lambda u):x\in N, u\in\R\}$ and obtain
$\Fl{t}{Y}(x,u,\lambda u)= \begin{pmatrix} x\\K_x(t,u)\\ -t+\lambda K_x(t,u) \end{pmatrix}$, denoting this restriction by
$\Phi\colon \R\times S \rightarrow M$. 

The next step is to determine $\Phi^*l=:g$ and the isometry $\Phi\colon (\R\times S, g) \rightarrow (M,l)$ explicitly
in order to describe the metric splitting in detail. We summarize the statement from \cite[Thm.\ 1.1]{BS:05} (see also
\cite[Thm.\ 3.78]{MS:08}).

\begin{thm}[Bernal, S\'{a}nchez]\label{thm-bs}
 Let $(P,r)$ be a globally hyperbolic Lorentzian manifold, then $(P,r)$ is isometric to $(\R\times S, -\beta dt^2 +
r_t)$, where $S$ is a smooth spacelike Cauchy hypersurface, $t\colon\R\times S\rightarrow \R$ is the projection onto the
first
factor, $\beta\colon\R\times S\rightarrow (0,\infty)$ a smooth function, and $r_t$ is a Riemannian metric on each
$S_t:=\{t\}\times S$, which varies smoothly with $t$.
\end{thm}

The metric splitting has important applications in general relativity -- for example in the initial value problem for the
Einstein equation and for the solution theory of the wave equation.

\bigskip
Upon identification of $S$ with $N\times \R$ via $(x,u,\lambda u) \leftrightarrow (x,u)$ we arrive at the following theorem.

\begin{thm} (Metric splitting) \label{thm-ms}
 With the notation used above we have 
\begin{equation*}
g_{(t,x,u)}=\frac{-1}{2\lambda - A(t,x,u)} dt^2 + H_t(x,u) \quad
  \text{for $(t,x,u)\in \R\times N\times \R$}\ ,
\end{equation*}
where $A(t,x,u):=a(x,K_x(t,u))>0$, $H_t(x,u)$ is the Riemannian metric on $\{t\}\times N\times\R$ for every $t\in\R$ given locally by
\begin{align*}
 H_t(x,u) &= \ \sum_{i,j=1}^n{\left(h_{ij}+(2\lambda-A)\pder{K_x}{x^i}\pder{K_x}{x^j}\right)dx^i dx^j}\\ 
	  &+ 2(2\lambda-a)\sum_{i=1}^n{\pder{K_x}{x^i}\ dx^i du} + \frac{(2\lambda-a)^2}{2\lambda-A}\ du^2\ .
\end{align*}
\end{thm}
\begin{pr}
 Let $(t,x,u)\in \R\times N \times \R$ and let $x^1,\ldots, x^n$ be a coordinate system on $N$. Then we calculate
 $g_{(t,x,u)}=\Phi^*l_{(t,x,u)}$ using the corresponding matrices
{\onehalfspacing
 \begin{equation*}
 l_{\Phi(t,x,u)} = \begin{pmatrix} h_x 	& 0  	& 0\\
				   0	& -A	& 1\\
				   0	& 1 	& 0 \end{pmatrix},\ 
 T_{(t,x,u)}\Phi = \begin{pmatrix} I_n  	& 0		& 0\\
 				 \nabla_x K^\intercal		& \pder{K}{t} 		 & \pder{K}{u}\\
 				 \lambda \nabla_x K^\intercal	& -1+\lambda \pder{K}{t} & \lambda \pder{K}{u}
\end{pmatrix}
 \end{equation*}}(note that we put the Riemannian part in the upper left corner in concordance with \ref{eq-l-mat}, then the
$t$ and $u$ part). 
 
We obtain
{\onehalfspacing
 \begin{equation*}
g_{(t,x,u)}= \begin{pmatrix} 	h_x + (2\lambda- A)\nabla_x K \nabla_x K^\intercal   	& 0			&
(2\lambda-a)\nabla_x K\\
				0						& \frac{-1}{2\lambda -A}& 0\\
 				(2\lambda-a)\nabla_x K^\intercal		& 0 			&
\frac{(2\lambda-a)^2}{2\lambda-A} \end{pmatrix} \ ,
\end{equation*}}where we have used that $\pder{F_x}{z}\pder{K}{t}=1$, $\pder{F_x}{z}\pder{K}{u}=2\lambda - a$ and
$\pder{F_x}{z}(x,K_x(t,u))
= 2\lambda -A(x,t,u)$. Observe that $H_t(x,u)$ is positive definite, since by setting $y:=\tilde{v} \nabla_x K$,
$w:=v_{n+1} \in \R$ for $v^\intercal:=(\tilde{v},v_{n+1})\in\R^{n+1}$, we obtain
\begin{equation}\label{eq-H-est}
 v^\intercal H_t v =  \underbrace{\tilde{v}^\intercal h_x \tilde{v}}_{\geq 0} +
\underbrace{\frac{1}{2\lambda-A}}_{>0} \bigl( (2\lambda-A)y + (2\lambda-a)w \bigr) ^2 \geq 0 \ .
\end{equation}
Therefore $H_t$ is a Riemannian metric on each $\{t\}\times S$; in conclusion, we obtained the metric splitting as in
\cite[Thm.\ 1.1]{BS:05}.
\end{pr}

In summary, $(M,l)$ is isometric to $(\R\times S, g)$ with
\begin{equation}\label{eq-ms-ms}
 g_{(t,x,u)} = -\theta(t,x,u) dt^2 + H_t(x,u)\ ,
\end{equation}
where 
\begin{equation}\label{eq-th-est}
0 < \frac{1}{2\lambda}\leq \theta(t,x,u)= \frac{1}{2\lambda-A(t,x,u)} = \frac{1}{2\lambda - a(x,K_x(t,u))}\leq
\frac{1}{\lambda}\ .
\end{equation}

\section{Metric splitting for non-smooth NPWs}\label{sec-nsm}
\subsection*{Generalized metric}
\addcontentsline{toc}{subsection}{Generalized metric}
Now we want to allow for a non-smooth profile function $a$, while the Riemannian metric $h$ on $N$ still is smooth.
Technically we will view $a$ as a generalized function in the sense of Colombeau and represent it by a net of smooth
functions $(a_\ep)_\ep$. If one is interested in solving the (vacuum) Einstein equations then the regularity of $a$ improves.
The Einstein vacuum equations for NPWs are
\begin{align*}
 \Delta_x a(x,u) = 0\ ,\\
 \Ric^N = 0\text{ on }N\ 
\end{align*}
(cf.\ \cite[p.\ 85]{FS:06}), where $\Delta_x$ is the Laplace operator on $(N,h)$. Solutions $a$ of the Laplace equation are
always analytical in $x$ for every $u\in\R$ and therefore $a$ can be non-smooth only with respect to $u$. 

Let $I:=(0,1]$, $a_\ep\in\Cinf(N\times\R)$ with $a_\ep\geq 0$ and $\|a_\ep\|_\infty < \lambda$ for all $\ep\in I$. We
assume that $(a_\ep)_\ep$ is moderate and  smooth with respect to $\ep$, thus   defines a class $a:=[(a_\ep)_\ep]$ in $\G(N\times\R)$.
(Note that smoothness in $\ep$ can be weakened to continuity by \cite[Thm.\ 3.9]{BK:12}).

\begin{lem}\label{lem-gen-met}
 With the notation used above let $l_\ep:=\pi^*(h) + 2 du dv - a_\ep(.,u)du^2$ and 
set $l:=[(l_\ep)_\ep]\in \G_2^0(M)$.  Then $l$ is a generalized Lorentzian metric and so $(M,l)$ is a generalized Lorentzian
manifold (\cite[Def.\ 3.4]{KS:02b}).
\end{lem}
\begin{pr}
 We observe that $\det((l_\ep)_{(x,u,v)})=-\det(h_x)<0$ for all $\ep\in I$ and $x\in N$ and by \eqref{eq-cc-ev-gen-case}
for $a_\ep$ we conclude that the eigenvalues depending on $\ep$, namely, $\mu_1^\ep, \mu_2^\ep$ of $(l_\ep)_{(x,u,v)}$
satisfy the following estimates:
\begin{equation*}
 \mu_2^\ep < -1\quad \text{ and } \quad \mu_1^\ep > \frac{-\lambda}{2} + \sqrt{\frac{\lambda^2}{2}+1}>0\qquad \forall \ep\in
I\ .
\end{equation*}
Since the other $n$ eigenvalues of $(l_\ep)_{(x,u,v)}$ are the positive eigenvalues of $h_x$,the index of $l$
is $1$ and hence by \cite[Thm.\ 3.1 and Prop.\ 3.3]{KS:02b} $l$ is a generalized Lorentzian metric.
\end{pr}

\subsection*{The generalized metric splitting}
\addcontentsline{toc}{subsection}{The generalized metric splitting}
From now on we assume in case $\dim(N)=n>0$ that the trajectories of
\begin{equation*}
D^N_{\dot \xi}\dot \xi = \frac{1}{2} \nabla_x a_\ep(\xi, \alpha)
\end{equation*}
are complete for every $\alpha\in\R$ and all $\ep>0$ small. Then we are able to apply Theorem \ref{thm-flow-sm} for every $\ep$ and deduce that $S^k_\lambda$ is a Cauchy hypersurface for every $l_\ep$ ($\ep>0$ small) for any $k \in \R$. Consequently, uniform
bounds on $\pder{a_\ep}{u}$ with respect to $\ep$ will be sufficient (cf.\ \cite[Prop.\ 2]{CRS:12}) for the completeness
of these trajectories.

As in the construction in Theorem \ref{thm-flow-sm} we define $F\colon I\times N\times \R \rightarrow \R$ by 
$F_\ep(x,z):=F(\ep,x,z):= \int_0^z(2\lambda - a_\ep(x,s))ds$, which depends smoothly on all variables by assumption and
the net $(F_\ep)_\ep$ is moderate. Then as in the smooth case $F_{\ep,x}:=F(\ep,x,\ .\ )\colon\R\rightarrow\R$ is smooth and
bijective for all $\ep\in I, x\in N$. So for fixed $x\in N$, $\ep\in I$ we can define
$K_\ep(x,t,u):=K(\ep,x,t,u):=F_{\ep,x}^{-1}(t +
F(\ep,x,u))$, which is globally defined. By the implicit function theorem, applied to 
\begin{equation} \label{eq-impl-fct}
F(\ep,x,K(\ep,x,t,u)) = t + F(\ep,x,u)\ ,
\end{equation}
we conclude that $K$ is smooth. Moreover, $F_\ep(K\times L)\subseteq [\lambda \min(L),  2 \lambda \max(L)]$ is compact for any  $K\subcomp N$, $L\subcomp \R$, hence the class $[(F_\ep)_\ep]$ is c-bounded. From \eqref{eq-impl-fct}, the
definition of $F$, and by $\lambda \leq 2\lambda - a_\ep(x,s) \leq 2 \lambda$ for all $x\in N, s\in\R, \ep\in I$ we get that
\begin{equation*}
 \lambda\ K(\ep,x,t,u) \leq t + F(\ep, x, u) \leq 2\lambda\ K(\ep,x,t,u)\ ,
\end{equation*}
or, 
\begin{equation} \label{eq-K-c-bou}
 \frac{1}{2 \lambda}(t + F(\ep, x, u)) \leq K(\ep,x,t,u) \leq \frac{1}{\lambda}(t + F(\ep,x,u))\ .
\end{equation}
Now by \eqref{eq-K-c-bou} it is obvious that $(K_\ep)_\ep$ is also c-bounded. To show that $(K_\ep)_\ep$ is moderate, it
suffices to observe that \eqref{eq-impl-fct} yields
\begin{equation}\label{eq-pder-K}
\pder{K_\ep}{x}(x,t,u) = \pder{F_\ep}{x} (\pder{F_\ep}{z})^{-1} = \frac{-1}{2\lambda -a_\ep(x,K_\ep(x,t,u))}
\int_0^{u}\pder{a_\ep}{x}(x,s) ds\ .
\end{equation}

\bigskip
As in Subsection \ref{sub-ms} let $S:=S^0_\lambda=\tla(\{0\}) = \{(x,u,\lambda u):x\in N, u\in\R\}$ and define $\Phi\colon I
\times \R \times S \rightarrow M$ by $\Phi_\ep(t,x,u):= \Phi(\ep,t,x,u) := 
\begin{pmatrix}
x \\
K_\ep(x,t,u)\\
-t+\lambda K_\ep(x,t,u) 
\end{pmatrix}$. Then $\Phi_\ep$ is a diffeomorphism for every $\ep\in I$ (by Theorems \ref{thm-flow-sm} and \ref{thm-ms} for
fixed $\ep\in I$) and since its components are c-bounded, it is c-bounded and hence
$\Phi:=[(\Phi_\ep)_\ep]\in\G[\R\times S,M]$. (It is clear that $\Phi_\ep$ is the flow of
$Y_\ep:=\frac{\grad(\tla)}{l_\ep(\grad(\tla),\grad(\tla))}$ restricted to $S$, where the gradient is with respect to
$l_\ep$.)

\bigskip
To develop here a more general version of a globally hyperbolic metric splitting we will employ the concept of a generalized diffeomorphism, a first variant of which was introduced in \cite[Section 4]{KS:99} and later consolidated in the unpublished theses \cite[Def.\ 5.35]{Steinbauer:00}. For convenience of the reader we state here the definition in full detail (the adaptation to manifolds in place of open sets is straightforward.)
\begin{defi}(Generalized diffeomorphism) \label{def-gen-diff}
 Let $\Omega\subseteq \R^n$ be open. We call $T\in\G(\Omega,\R^n)$ a \underline{generalized diffeomorphism} if there exists
$\eta>0$ such that
\begin{enumerate}
 \item \label{def-gen-diff-1} There exists a representative $(t_\ep)_\ep$ of $T$ such that $t_\ep\colon\Omega\rightarrow
t_\ep(\Omega)=:\tilde{\Omega}_\ep$ is a diffeomorphism for all $\ep\leq \eta$ and there exists $\tilde{\Omega}\subseteq\R^n$
open with $\tilde{\Omega}\subseteq \bigcap_{\ep\leq\eta}\tilde{\Omega}_\ep$.
 \item The inverses $(t_\ep^{-1})_\ep$ are moderate, i.e., $(t_\ep^{-1})_\ep\in\EM(\tilde{\Omega},\R^n)$ and there exists
$\Omega_1\subseteq\R^n$ open, $\Omega_1\subseteq\bigcap_{\ep\leq\eta}t_\ep^{-1}(\tilde{\Omega})$.
 \item Setting $T^{-1}:=[(t_\ep^{-1}|_{\tilde{\Omega}})_\ep]$, the compositions $T\circ T^{-1}$ and $T^{-1}\circ
T|_{\Omega_1}$ are elements of $\G(\tilde{\Omega},\R^n)$ respectively $\G(\Omega_1, \R^n)$. (It is then clear that $T\circ
T^{-1} = id_{\tilde{\Omega}}$ and $T^{-1}\circ T|_{\Omega_1} = id_{\Omega_1})$.
\end{enumerate}
\end{defi}

We are now in a position to show that our collection of diffeomorphisms $(\Phi_\ep)_\ep$ represents a generalized
diffeomorphism in the sense of Definition \ref{def-gen-diff}.
\begin{prop}
 The generalized function $\Phi$ is a generalized diffeomorphism, moreover $\Phi$ and its inverse
$\Phi^{-1}$ are c-bounded, hence $\Phi\in\G[\R\times S, M]$ and $\Phi^{-1}\in\G[M,\R\times S]$.
\end{prop}
\begin{pr}
We observe that $\Phi_\ep(\R\times S)=M$ for all $\ep\in I$, so the first point of Definition
\ref{def-gen-diff} is clearly satisfied and similarly for $\Psi_\ep:=\Phi_\ep^{-1}$ we have that the
image $\Psi_\ep(M)=\R\times S$ has no dependence on $\ep$. From the construction in \cite{BS:05} we know that for each
$\ep\in I$ the diffeomorphism $\Psi_\ep$ is given as $\Psi_\ep = (\tau_\lambda, \Pi_\ep)$, where $\tau_\lambda$ is the time
function and $\Pi_\ep(p)$ is the unique intersection point of the flowline of $Y_\ep$ starting at $p\in M$ with $S$.
Therefore it suffices to show that $\Pi_\ep$ is moderate respectively c-bounded to show the corresponding property for
$\Psi_\ep$.

From Theorem \ref{thm-flow-sm} we know that if we start at $p:=(x,u,v)\in M$ then $\Fl{v-\lambda u}{Y_\ep}(p)\in S$, hence
\begin{equation}\label{eq-pi_ep}
 \pi_\ep(p)=\begin{pmatrix}
             x\\
	     K_\ep(x,v-\lambda u, u)\\
	     \lambda K_\ep(x,v-\lambda u, u)
            \end{pmatrix}\ ,
\end{equation}
which is clearly moderate and c-bounded since $(K_\ep)_\ep$ is.
\end{pr}

In an attempt to generalize the notion of global hyperbolicity to generalized Lorentzian manifolds the concept of a
so-called \emph{globally hyperbolic metric splitting} has been introduced in \cite[Def.\ 6.1]{HKS:12}. Our investigations here have shown the need to adapt this definition  to also allow for a generalized diffeomorphism instead of a classical one.
\begin{defi}(Globally hyperbolic metric splitting)\label{def-gms-ghms}
 Let $g$ be a generalized Lorentz metric on the smooth $(n + 1)$-dimensional manifold M. We say that $(M, g)$ allows a
\underline{globally hyperbolic metric splitting} if there exists a generalized diffeomorphism $\Phi \colon M \rightarrow
\R\times
S$, where $S$ is an $n$-dimensional smooth manifold such that the following holds for the pushed forward generalized Lorentz
metric $\lambda := \Phi_*g$ on $\R\times S$:
\begin{enumerate}[(a)]
  \item There is a representative $(\lambda_\ep)_\ep$ of $\lambda$ such that every $\lambda_\ep$ is a Lorentz metric and each
slice $\{t_0\}\times S$ with arbitrary $t_0\in\R$ is a (smooth, spacelike) Cauchy hypersurface for every $\lambda_\ep$.
  \item \label{def-gms-ghms-b} We have the metric splitting of $\lambda$ in the form
    \begin{equation*}
     \lambda=-\theta dt^2 + H\ ,
    \end{equation*}
where $H\in\Gamma_\G(\mathrm{pr}_2^*(T_2^0 S))$ is a $t$-dependent generalized Riemannian metric and $\theta\in\G(\R\times
S)$ is globally bounded and locally uniformly positive, i.e., for some (hence any) representative $(\theta_\ep)_\ep$ of
$\theta$ and for every $K \subcomp \R\times S$ we can find a constant $C > 0$ such that $\theta_\ep(t,x) \geq C$ holds
for small $\ep>0$ and $x \in K$.
  \item \label{def-gms-ghms-c} For every $T > 0$ there exists a representative $(H_\ep)_\ep$ of $H$ and a smooth complete
Riemannian metric $\rho$ on $S$ which uniformly bounds $H$ from below in the following sense: for all $t\in [-T, T]$, $x\in
S$, $v \in T_x S$, and $\ep\in I$
\begin{equation*}
 (H_\ep)_t(v,v)\geq \rho(v,v)\ .
\end{equation*}
\end{enumerate} 
\end{defi}

The hypothesis of the following corollary is satisfied if, e.g.\ $\nabla_x a_\ep$ is locally bounded, uniformly with respect to $\ep$, as can be seen from \eqref{eq-pder-K}.
\begin{cor}\label{cor-gms}
If $\nabla_x K_\ep$ is locally bounded, uniformly with respect to $\ep$, then the generalized Lorentzian manifold $(M,g)$
allows a globally hyperbolic metric splitting in the form of \eqref{eq-ms-ms} where $\theta\in\G(\R\times S)$ and
$H_t\in\Gamma_\G(\mathrm{pr}^*_2(T_2^0 S))$ satisfy the
second and third part of Definition \ref{def-gms-ghms}.
\end{cor}
\begin{pr}
We show that the $\ep$-wise constructions are designed in such a way that the global metric splitting can be carried out in the generalized sense.
 \begin{enumerate}[(a)]
  \item 
    This is clear from the construction.
  \item
    The metric splitting in this form was given before and from \eqref{eq-th-est} we see that $\theta$ is globally
bounded and globally uniformly positive.
  \item
    First we construct $\rho$ locally and then we extend it to a globally defined Riemannian metric. So let $T>0$ and fix
$x\in N$, $u\in \R$ and $\ep\in I$, then for $v\in\R^{n+1}\cong T_{(x,u)}(N\times\R)$ of the form $v^\intercal =
(\tilde{v}^\intercal,w)$, we get from \eqref{eq-H-est} that 
\begin{equation}\label{eq-H_t-ineq}
 (H_\ep)_t(v,v) \geq \tilde{v}^\intercal h_x \tilde{v} + \lambda \bigl( \tilde{v}^\intercal \nabla_x K_\ep +
\frac{1}{2}w \bigr)^2 \ .
\end{equation}
  Now we set $d:= \sup_{\ep\in I, t\in [-T,T]}|\nabla_x K_\ep(x,t,u)| < \infty$ and since $h_x$ is a positive definite
metric on $\R^n$ there is a constant $\alpha>0$ such that $h_x(\tilde{v},\tilde{v})\geq \alpha |\tilde{v}|^2$ for all
$\tilde{v}\in\R^n$, i.e., $h_x$ can be bounded from below by a multiple of the euclidean norm. At this point we consider
the case that $|w|\leq 4 d |\tilde{v}|$:
\begin{align*}
 (H_t)_\ep(\tilde{v},\tilde{v}) &\geq \alpha |\tilde{v}|^2 = \frac{\alpha}{2}|\tilde{v}|^2 + \frac{\alpha}{2}|\tilde{v}|^2
\geq
 \frac{\alpha}{2}|\tilde{v}|^2 + \frac{\alpha}{16 d^2} |w|^2\\
 &\geq \min(\frac{\alpha}{2},\frac{\alpha}{16 d^2}) \bigl( |\tilde{v}|^2 + w^2 \bigr) \ ,
\end{align*}
where we used that the second term in \eqref{eq-H_t-ineq} is non-negative and the bounds on $h_x$ and $|\tilde{v}|$ from
below. Now the case where $|w|>4 d |\tilde{v}|$: 
to this end we estimate $|\tilde{v}^\intercal \nabla_x K_\ep + \frac{1}{2}w| \geq \frac{1}{2} |w| - |\tilde{v}^\intercal
\nabla_x K_\ep| \geq \frac{1}{2}|w| - |\tilde{v}||\nabla_x K_\ep| \geq \frac{1}{2}|w| - \frac{1}{4}
|w| = \frac{1}{4}|w|$. This allows us to estimate
\begin{align*}
 (H_t)_\ep(v,v)\geq \alpha |v|^2 + \lambda \frac{1}{16}w^2 \geq \min(\alpha,\frac{\lambda}{16})\bigl(|\tilde{v}|^2 + w^2
\bigr)\ ,
\end{align*}
where we used the estimate given above, the bounds on $h_x$, $\nabla_x K_\ep$ and the Cauchy-Schwarz inequality. In summary we
see that 
\begin{equation*}
(H_t)_\ep(v,v)\geq \min(\frac{\alpha}{2}, \frac{\alpha}{16 d^2}, \frac{\lambda}{16}) |v|^2\ ,
\end{equation*}
for all $\ep\in I$, $t\in[-T,T]$ and $v\in\R^{n+1}$.

To get a globally defined Riemannian metric on $N\times \R$ which is a lower bound of $(H_t)_\ep$ from below (uniformly in
$\ep$), employ a partition of unity on $N\times\R$ to pass from local to global constructions.

 \end{enumerate}
\end{pr}

\begin{rem}
 If we drop the boundedness assumption on $\nabla_x K_\ep$ in the previous corollary, then a lower bound in terms of a Riemannian metric need not exist. It already fails at a point: Fix $(x,u)\in N\times \R$, $t\in[-T,T]$ and assume that
$\limsup_{\ep\searrow0}|\nabla_x K_\ep(x,t,u)|=\infty$. Assume that we would have a lower bound:
$(H_t)_\ep(v,v)\geqq c |v|^2$ for all $\ep\in I$, $v\in\R^{n+1}$. Set $\tilde{v}:=\nabla_x K_\ep$ and
$w:=-\frac{2\lambda-A}{2\lambda-a}\tilde{v}^\intercal \nabla_x K_\ep$ in \eqref{eq-H-est}, then $O(|\nabla_x K_\ep|^4) = c
|v|^2 \leq \nabla_x K_\ep^\intercal h_x \nabla_x K_\ep = O(|\nabla_x K_\ep|^2)$, a contradiction.
\end{rem}

\section{Approximation and limits}\label{sec-app-app}
In the final part we apply the results of Section \ref{sec-nsm} to specific classes of non-smooth Lorentzian manifolds. Let $a_0\in L^\infty(N\times\R)$ with $a_0\geq 0$ almost everywhere and $\la > 0$ such that $\|a_0\|_{L^\infty}<\lambda$. Then as in
Theorem \ref{thm-flow-sm} the function $F_{0,x}(u):=\int_0^u (2\lambda - a_0(x,s))ds$ can be defined for every $x\in N$ and it
is again a bi-Lipschitz homeomorphism for every $x\in N$. This allows us to define
$K_{0,x}(t,u):=F_{0,x}^{-1}(t+F_{0,x}(u))$, which is also Lipschitz continuous for every $x\in N$. Without further
assumptions on $a_0$, the function $F_0 \colon N\times \R\rightarrow \R$ need not even be continuous (e.g., $N:=\R$ and
$a_0(x,u):=H(x)$ the Heaviside function). Assume that there is a function $c\colon\R\rightarrow(0,\infty)$ that is locally
integrable satisfying
\begin{equation}\label{eq-a0-Lip}
 |a_0(x,s)-a_0(y,s)|\leq c(s)\, d^h(x,y)\qquad (\forall x,y\in N, s\in \R)\ ,
\end{equation}
i.e., $a_0(.,s)$ is Lipschitz continuous for all $s\in\R$. Then $F_0$ is locally Lipschitz continuous on $N\times\R$:
First we prove that $F_0$ is separately Lipschitz continuous with uniform Lipschitz constant in $x$ and as observed above
$F_0$ is Lipschitz continuous in $s$ uniformly in $x$. As for Lipschitz continuity with respect to $x$ fix $s\in\R$, let
$x\in N$ and $U$ a neighborhood of $x$, then for all $x_1,x_2\in U$:
\begin{equation*}
 |F_0(x_1,s)-F_0(x_2,s)| \leq \int_0^s{|a_0(x_1,\tau)-a_0(x_2,\tau)|d\tau} \leq d^h(x_1,x_2)\int_0^s c(\tau)d\tau\ .
\end{equation*}
Consequently for $a,b\in\R$ with $s\in (a,b)$ and $C:=\int_0^b{c(\tau)d\tau}$ ($<\infty$, since $C$ is locally integrable)
$F_0$ is Lipschitz continuous on $U\times(a,b)$ with Lipschitz constant $\max(C,2\lambda)$: For all $s_1,s_2\in
(a,b)$, $x_1,x_2\in U$
\begin{align*}
|F_0(x_1,s_1)-F_0(x_2,s_2)|&\leq |F_0(x_1,s_1)-F_0(x_1,s_2)| + |F_0(x_1,s_2)-F_0(x_2,s_2)|\\&\leq 2\lambda |s_1-s_2| + C
d^h(x_1,x_2)\ . 
\end{align*}

At this point we will employ a variant of the implicit function theorem for Lipschitz continuous functions, the proof of which is routine: Let $U \subseteq \R^m$, $V \subseteq \R^n$ be open, $a \in U$, $b \in V$, and $F \col U \times V \to \R^n$ be Lipschitz continuous. Suppose that $F(a,b) = 0$ and there exists $L > 0$ such that
$$
    |F(x,y_1) - F(x,y_2)| \geq L \, |y_1 - y_2|
$$   
holds for $(x,y_1)$ and $(x,y_2)$ near $(a,b)$. Then there is an open neighborhood $\widetilde{U}$ of $a$ and a Lipschitz continuous function $\vphi\col \widetilde{U} \to V$ such that $F(x, \vphi(x)) = 0$ for all $x \in \widetilde{U}$. 

The above statement allows us to conclude as in Section \ref{sec-nsm} that $F_0$ is a local bi-Lipschitz homeomorphism and
thus $K_0\colon N\times \R^2\rightarrow\R$ and $\Phi_0\colon\R\times S \rightarrow M$ are locally
Lipschitz continuous. Furthermore defining $\Psi_0\colon M\rightarrow\R\times S$,
by $\Psi_0=(\tau_\lambda, \Pi_0)$, where $\Pi_0$ is given by \eqref{eq-pi_ep} in terms of $K_0$, then $\Psi_0$ is also
locally Lipschitz continuous. Moreover $\Psi_0$ is the inverse of $\Phi_0$, thus $\Phi_0$ is a local bi-Lipschitz
homeomorphism.

In summary, even in this non-smooth setting we obtain a kind of topological splitting $M\cong \R\times S$, in fact
via a local bi-Lipschitz homeomorphism. However, at the moment the meaning of $\Phi_0$  in the context of Lorentzian
geometry is unclear. On the other hand we will see in the following that $\Phi_0$ is the limit of reasonable smooth
approximations (i.e., diffeomorphisms $(\Phi_\ep)_\ep$ as in Section \ref{sec-nsm}).

\bigskip
Now we assume that we are given $a_0\in L^\infty(N\times\R)$ with $a_0\geq 0$ almost everywhere and $|a_0|<\lambda$
as above but not necessarily with the Lipschitz continuity in the first argument. Moreover let $(a_\ep)_\ep$ be a net in
$\Cinf(N\times\R)$ satisfying $0\leq a_\ep < \lambda$ for all $\ep\in I$ and $a_\ep \to a_0$ (for $\ep\searrow 0$) in
$L_\loc^1(N\times\R)$ (based on a Lebesgue measure on the manifold $N\times\R$ in the sense of Dieudonn\'e \cite[Section
16.22]{Dieu:72:3}). Then $F_\ep\to F_0$ (for $\ep\searrow 0$) in $L_\loc^1(N\times\R)$: Let $K\Subset N$, $a,b\in\R$
with $a<b$, then 
\begin{equation*}
\bigl | \int_{K\times[a,b]}(F_\ep-F_0)\bigr | \leq 
\int_a^b \underbrace{\int_K \int_0^z|a_\ep(x,s)-a_0(x,s)|ds\ dx}_{\to 0 \text{ by }a_\ep\to a_0\text{ in }L_\loc^1(N\times\R)}\ dz\ .
\end{equation*}

Set $G_\ep(x,z):=(F_\ep)_x^{-1}(z)$ for $(x,z)\in N\times\R$ and $\ep\in [0,1]$, then $G_\ep\to G_0$ (for $\ep\searrow
0$) in $L_\loc^1(N\times\R)$: The defining equation for $G_\ep$ is $F_\ep(x,G_\ep(x,z))=z$ for all $(x,z)\in N\times\R$.
Therefore $\frac{1}{2\lambda}\leq \pder{G_\ep}{z}\leq\frac{1}{\lambda}$ and consequently $|G_\ep(x,z)-G_\ep(x,0)|\leq
\int_0^z|\pder{G_\ep}{z}(x,s)|ds\leq\frac{|z|}{\lambda}$. Thus $(G_\ep(.,z))_\ep$ is bounded for every $z\in\R$, hence
$(G_\ep)_\ep$ is bounded on compacta in $z$. Applying the theorem of Arzel\`{a}--Ascoli to the net $(G_\ep(x,.))_\ep$ of
equicontinuous smooth functions (their Lipschitz constants are bounded by $\frac{1}{\lambda}$) for every $x\in N$, yields a
uniformly convergent subsequence $(G_{\ep_k}(x,.))_k$ in $\Con^0([a,b])$ for suitable $a,b\in\R$. Denote this limit by
$g(x,.)\in\Con^0([a,b])$, then by the defining equation of $G_\ep$ and the equicontinuity of $(G_\ep)_\ep$ we conclude
that $G_{\ep_k}\to G_0$ in $L^1_\loc(N\times\R)$: For fixed $x\in N$ the net $F_\ep(x,G_\ep(x,.))$ converges uniformly on
$[a,b]$ to $F_0(x,g(x,.))$, which has to be $\textrm{id}_{[a,b]}$, hence $G_0 = g$ on $N\times[a,b]$. This yields the
$L_\loc^1$-convergence of $G_\ep$ to $G_0$. In summary all subsequences of $(G_\ep)_\ep$ have to converge to $G_0$,
hence $G_0$ is the limit of $(G_\ep)_\ep$.

Similarly one shows that $K_\ep\to K_0$ in $L^1_\loc(N\times\R^2)$: for $(x,t,u)\in N\times\R^2$ we obtain
\begin{equation*}
 |K_\ep(x,t,u)-\underbrace{K_\ep(x,t,0)}_{= G_\ep(x,t)}| \leq 2 |u| + |G_\ep(x,t)|\ , 
\end{equation*}
which is bounded on compacta in $t$ and $u$. Then continuing as before using the defining equation of $K_\ep$, i.e.,
$F_\ep(x,K_\ep(x,t,u)) = t + F_\ep(x,u)$, and the Arzel\`{a}--Ascoli theorem we conclude that all subsequences of
$(K_\ep)_\ep$ converge to $K_0$ in $L_\loc^1(N\times\R^2)$, thus the limit of $(K_\ep)_\ep$ is $K_0$. All in all we establish
that $(\Phi_\ep)_\ep$ converges to $\Phi_0$ and $(\Psi_\ep)_\ep$ converges to $\Psi_0$ in $L_\loc^1(\R\times S)$
respectively $L_\loc^1(M)$ ($\Phi_\ep(t,x,u)$ is just a linear combination of $K_\ep(x,t,u)$, $x$ and $t$ respectively
for the convergence of $(\Psi_\ep)_\ep$ one uses additionally the uniform converge of $(K_\ep(x, ., .))_\ep$). 

\bigskip
In the two-dimensional case, i.e., $M=\R^2$, one can obtain stronger results:
assuming $a_\ep \to a_0$ in $L_\loc^1(\R)$ we obtain $\Phi_\ep\to\Phi_0$ in $\Con^{0,\alpha}(L, \R^2)$ for every
$0<\alpha<1$ and $L\Subset \R^2$, where $\Con^{0,\alpha}(L, \R^2)$ are the $\alpha$-H\"older continuous functions from $L$
to $\R^2$.

\bigskip
In summary, we see that if $a_0$ satisfies the Lipschitz condition \eqref{eq-a0-Lip} we get the topological splitting $M\cong
\R\times S$ via a local bi-Lipschitz homeomorphism and assuming that $a_\ep\to a_0$ in $L_\loc^1(N\times\R)$ the
corresponding diffeomorphisms converge to this homeomorphism. Thus by approximating the non-smooth profile function $a_0$
this procedure yields an approximation of the metric splitting. 
Convoluting $a_0$ by an appropriate mollifier $\rho_\ep$ $(\ep\in (0,1])$ yields the $L_\loc^1(N\times\R)$ convergence of
$a_\ep:=a_0 * \rho_\ep$ to $a_0$.

\subsection*{Remark on the Cauchy problem for the wave equation}
Note that the generalized Lorentzian metric $[(l_\ep)_\ep]=l$ with representative given in Lemma \ref{lem-gen-met} is
\emph{weakly singular} (simple calculation and uniform bounds on $(a_\ep)_\ep$) in the sense of \cite[Subsection 5.1,
condition (A)]{HKS:12} and the gradient of $\tla$ is uniformly timelike by Proposition \ref{prop-tla}. Let us assume in addition that $\nabla a_\ep$ is bounded uniformly in $\ep$, then it is easily seen from \eqref{eq-pder-K} that $\nabla_x K_\ep$ is locally bounded, uniformly with respect to $\ep$ (as assumed in Corollary \ref{cor-gms}). Hence due to the structure of the generalized diffeomorphism $\Phi$, $(\R\times S,\lambda)$ satisfies the conditions of \cite[Thm.\ 6.3]{HKS:12}, which establishes global existence and uniqueness of generalized solutions to the Cauchy problem on $\R\times S$. Transporting this generalized solution back to $(M,l)$ via $\Phi$ yields a solution of the Cauchy problem of the wave equation on $M$:

\begin{equation*}
\left\{ \!\begin{aligned}
      \Box &u=0\text{ on }M\ ,\\
      &u=u_0\text{ on }S\ ,\\
      \nabla_{\hat{\xi}}&u = u_1 \text{ on }S\ .
\end{aligned} \right.
\end{equation*}

Here the unit normal vector field $\hat{\xi}$ of $S$ corresponds to $\frac{1}{\sqrt{\theta}}\partial_t$ on $\R\times S$, where
$\theta\in\G(\R\times S)$ is given by \eqref{eq-th-est} and $u_0, u_1\in \G(S)$ with compact supports.

\begin{rem}
Observe that in case of (distributional) convergence $a_\ep \to a_0$ (as $\ep \to 0$) the uniform boundedness of
$\|\nabla a_\ep\|$ implies that $a_0$ is (locally) Lipschitz continuous: In fact, in any coordinate neighborhood and for any
test function $\vphi$ supported there, we have $|\dis{\d_j a_\ep}{\vphi}| \leq \|\nabla a_\ep\|_{L^\infty} \cdot \| \vphi
\|_{L^1} \leq C \, \| \vphi \|_{L^1}$ with $C$ independent of $\ep$; hence $|\dis{\d_j a_0}{\vphi}|  \leq C \, \| \vphi
\|_{L^1}$, so that $\d_j a_0$ belongs to $(L^1)' = L^\infty$; hence finally $a_0$ is Lipschitz continuous by Rademacher's
Theorem. Therefore in this case we obtain a topological splitting as constructed in the beginning of the current section.
Moreover this allows us to conclude the following about generalized Lorentzian manifolds $(M,l)$ with $l$ as in Lemma
\ref{lem-gen-met}: If $M$ is not homeomorphic to $\R \times S$, via a local bi-Lipschitz homeomorphism, then $(M,l)$ cannot be
globally hyperbolic in the sense that there exists no globally hyperbolic metric splitting according to Definition
\ref{def-gms-ghms}. 
\end{rem}

\section{Summary}
We provided a case study for the possibility of a globally hyperbolic metric splitting for a class of non-smooth space-times. Along the way we have explicitly constructed the metric splitting for the smooth case and showed that its crucial properties can be transfered to the family of regularizations, namely giving a generalized diffeomorphism and splitting structure. Finally we proved convergence of the resulting regularizing metric splittings in case of bounded and or Lipschitz continuous Lorentz metrics.

\subsubsection*{Acknowledgment}
The authors thank the anonymous referees for suggesting several clarifications of details. This work was supported by project P25326 of the Austrian Science Fund.

\bibliographystyle{abbrv}
\bibliography{GloballyHyperbolicSplitting}
\addcontentsline{toc}{section}{References}

\texttt{Faculty of Mathematics, University of Vienna\\
Oskar-Morgenstern-Platz 1, A-1090 Vienna, Austria}\\
\texttt{guenther.hoermann@univie.ac.at}\\ \texttt{clemens.saemann@univie.ac.at}

\end{document}